# Spatial dependence in the rank-size distribution of cities

Rolf Bergs[1]

**Abstract**

Power law distributions characterise several natural and social phenomena. *Zipf's* law for cities is one of those. The study views the question of whether that global regularity is independent of different spatial distributions of cities. For that purpose, a typical *Zipf*ian rank-size distribution of cities is generated with random numbers. This distribution is then cast into different settings of spatial coordinates. For the estimation, the variables rank and size are supplemented by spatial spillover effects in a standard spatial econometric approach. Results suggest that distance and contiguity effects matter. This finding is further corroborated by three country analyses.

JEL-Classification: O18; R12

## 1. Introduction

*Zipf's* law of the rank-size distribution of cities is regarded as an enthrallment of rare social physics (Krugman 1997, 42-46). However, the inherent relationship between size and rank characterises a tautological relationship because size directly determines rank and *vice versa*. The independent variable is thus perhaps no true predictor but could just reflect a universal statistical phenomenon. A power law exponent typically close to -1 and a determination coefficient ($R^2$) close to one are an indication of that. Therefore some have questioned its relevance for economic analysis (notably Gan et al. 2006). However, such a ubiquitous regularity justifies the examination of the hidden factors behind. Various authors have done this (e.g. Gabaix 1999; Fujita et al. 1999; Brakman et al. 2009; Reggiani and Nijkamp 2015). In contrast to the frequency of words in languages and their rank distribution (Zipf 1949) the rank-size distribution of cities appears slightly more varied between countries (e.g. Rosen and Resnick 1980) and less stable over time (e.g. Brakman et al. 1999). Theoretically, it could be also affected by space via the contiguity and distance effects between cities of different size. Space, however, cannot predict to the rank-distribution of words.

To have a closer look at that context, I demonstrate a typical rank-size distribution of cities for a varied distribution of spatial coordinates. The objective is to see how much influence spatial contiguity or distance could have on the ranks of cities and thus the shape of the distribution. Different levels of concentration and dispersion of urban space among countries suggest such an exploration. Especially in countries with a geographical concentration of bigger cities there is reason to assume that these cities have evolved due to certain spatial advantages (e.g. raw materials, climate, accessibility or certain random determinants). Those city clusters are often characterised by specific industries of national importance. Whether and how dispersion and concentration forces determine the rank-size distribution of cities has been an object of research in urban economics.

Surprisingly, there has been little research shedding light on the spatial dimension of *Zipf's* law. Lalanne (2013) views the dichotomic urban structure of Canada. She rejects the *Zipf* law and the underlying scale invariance and shows that the Canadian urban structure has evolved in a deterministic process based on urban size (inhabitants within administratively defined boundaries), previous growth and the spatial setting. Coefficients for the years 1971 to 2001 vary between -0.77 and -0.81. The spatial component is not part of the *Pareto* regression; instead growth of cities is regressed on size and previous growth in standard spatial regression models (SEM and SAR). Rybski et al. (2013) simulate urban evolution under the assumption that urban growth takes place close to other urban settlements. They confirm *Zipf's* law and scale invariance of the clusters generated, except the primacy one. Le Gallo and Chasco (2007) explore *Zipf's* law for Spain by applying a SUR model which they cast into spatial autoregressive or error specifications. *Zipf's* law does not hold between 1900 and 2001. While the simple OLS estimate varies between -0.54 and -0.66, the extended spatial models deviate even further from *Zipf's* law, thus revealing spatial impacts. Cheng and Zhuang (2012), who look at urban evolution in central China under consideration of *Zipf's* law, show that the estimation of the *Pareto* coefficient has displayed an undulatory pattern between 1985 and 2009. They cannot confirm *Zipf's* law at any point of time. The OLS estimates are then augmented by spatial autoregressive or spatial error effects. Like in the study of Le Gallo and Chasco (2007), spatial dependence has a lowering impact on the original OLS *Pareto* estimate.[2]

---
[1] PRAC, Im Hopfengarten 19b, 65812 Bad Soden, Germany. E-Mail: RolfB@prac.de

In the study at hand I intend to show (i) whether and how spatial effects vary by concentration and dispersion and (ii) how they behave differently along the entire distribution of cities and specifically the upper *Pareto* tail. A major assumption is that the rank-size distribution is not homogeneously following a *Pareto* shape, but rather a combination of an upper *Pareto* and a lower lognormal section. The paper is not a predominant real world study but rather draws on a theory-led artificial simulation of *Zipf's* law. For this purpose, I random-generate a typical database of city rank-size distribution with both, an upper *Pareto* as well as a lower lognormally distributed tail. Here I assume a regularity, several authors have demonstrated empirically (a known controversy is: Malevergne et al. 2011 versus Eeckhout 2004). Different patterns of spatial distribution of cities are then explored. Finally, the simulation exercise is complemented by three country data sets of natural cities which had been detected by statistical segmentation of night satellite images.[3]

## 2. Methodological approach

The central assumption is as follows: For the typical upper *Pareto (3,1)* tail the expected exponent $\alpha$ is approximately 1 ($\pm 0.1$) as empirically confirmed ubiquitously. The relationship in terms of a cumulative distribution function is given by:

$$R = CS^{-\alpha}$$

or in its log-linear form:

$$\ln(R) = \ln(C) - \alpha \ln(S)$$

where R is rank, S means size and C is a constant. OLS or maximum likelihood are possible estimators of α. For the combined *Pareto* and lognormal tails the exponent does not fit *Zipf's* law as usual. *Zipf's* law is empirically confirmed only for the upper tail of bigger cities. It is however expected that, in accordance with the gravity model, interaction between two places (cities) depends on their mass *M* and their distance *D* so that interaction decreases with distance: $I_{j,k} = GM_j M_k D_{j,k}^{-\gamma}$.

G is a proportionality coefficient and γ an exponent showing the distance decay of interaction. This gravity relationship, known from international trade, also represents *Tobler's* law in its sense that "everything is related to everything else, but near things are more related than distant things…" (Chen 2015). Gravity is however not incorporated into *Zipf's* law, so that the *Pareto* exponent α≈1 in one country may remain stable under consideration of spatial interaction while, in another one, the incorporation of such interaction could perhaps lead to major deviations of α. This assumption suggests to further test spatial dependence in *Zipf's* law for cities by the following alternative

---

[2] It is to be noted that in the three studies on Canada, Spain and China size of cities is not defined by functionality but inhabitants within administratively defined boundaries of all cities (i.e. including the lower tail). This is a reason why *Zipf's* law does not hold in numerous cases (Brakman et al. 2009, 301-306).

[3] The term „Natural Cities" had been originally formulated from an architectural viewpoint by Alexander (1965) in a way how planned citied had evolved in contrast to natural ones. In the present debate, the term is used to capture the true functional expansion of a city independent of its administrative boundaries. For natural urban classification, novel databases such as nocturnal satellite imagery or social media data are used (Jiang and Miao 2014; Jiang et al. 2015; Bergs 2018)

standard spatial econometric procedures (spatial autoregressive versus spatial error model) where $D_{j,k}$ is included through a spatial weight matrix:

$$\ln(R) = \rho\omega\ln(R) + \ln(C) - \alpha\ln(S) + \varepsilon \quad \textbf{(SAR)}$$

or

$$\begin{cases} \ln(R) = \ln(C) - \alpha\ln(S) + \varepsilon & \textbf{(SEM)} \\ \varepsilon = \lambda\omega\varepsilon + v \end{cases}$$

where ω is a N x N row-standardised weight matrix to capture the distance decay effect and *C* is a constant while ρ and λ in addition to α are the coefficients estimated. The error term ε in the SEM case consists of spatial error and the residual *v*.

The right choice between both models can be determined by different tests, such as *Moran's I* of the residuals and (Robust) *Lagrange* multiplier (LM) tests. The different estimation types in Tables 1 and 2 correspond to the respective choice.

The simulated „countries" describe a distribution with in each case 109 observations for cities, the upper 50 being *Pareto* (3,1) distributed and the lower 59 with a lognormal shape. Both distributions are random-generated and then matched into one data set. This hybrid rank-size distribution is then combined with different distributions of coordinates X and Y, the first one being randomly generated to fit a normal distribution. Based on this configuration a spatial weight matrix is derived. With a normal distribution of both X and Y coordinates it is expected that spatial autocorrelation of city size is close to zero; big and small cities are thus spread evenly.

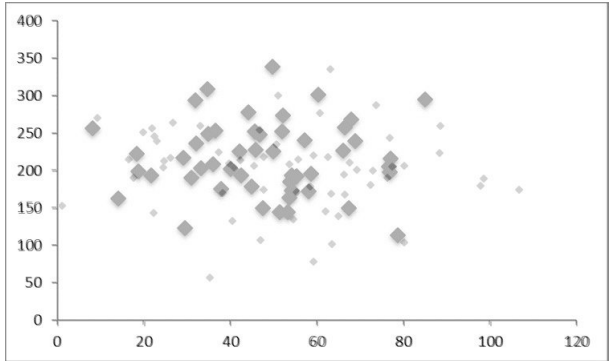

Fig. 1: Normal spatial distribution of cities

In a second variation with the same rank distribution the normally distributed coordinates of X and Y are both ranked as well, so that all cities are geographically positioned on a diagonal line, ordered along rank, the biggest city in the outer North-East, the smallest one in the outer South-West (like a one-dimensional *von-Thunen* assembly).

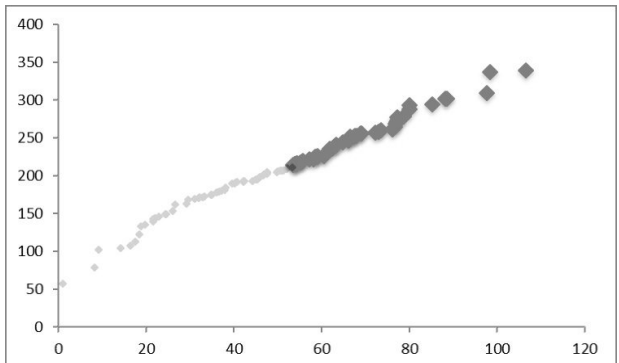

Fig. 2: Rank-size distribution of cities with ranked coordinates

It is expected that in this case both, rank as well as size, exhibit a high spatial autocorrelation of city size even though such an assembly is hardly encountered in the real world. In a third data set both, coordinates of the upper tail (*Pareto*) as well as the lower tail cities (lognormal) are normally distributed, however geographically separated and within a larger spatial frame, where coordinates of the upper tail are changed to 2X and 2Y. Here an indication of spatial dependence is likewise expected, but rather for the entire distribution.

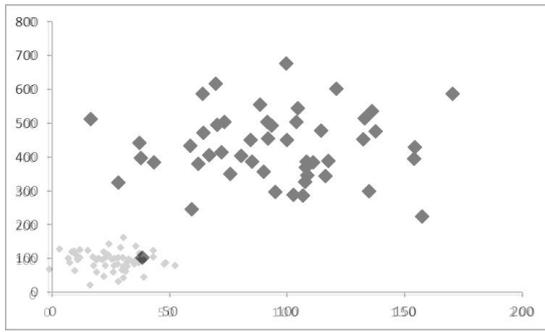
Fig. 3: Dichotomous spatial distribution of big and small cities

In a fourth variant there is a center-periphery pattern, however X coordinates of the bigger cities are again ranked so that the assembly follows a geographical distribution of biggest cities in the East, while the small cities remain normally distributed over space.

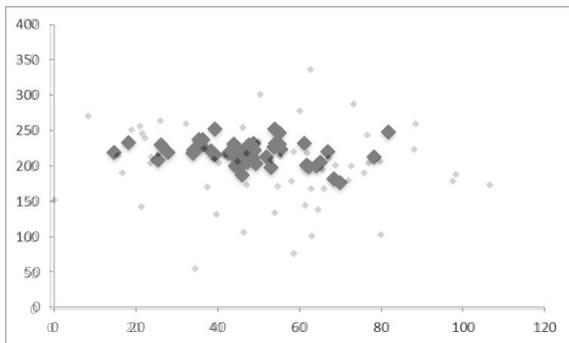
Fig.4: Non-normally distributed big cities and normally distributed small cities

Variants II, III and IV represent excessive non-normal formings and are hardly encountered in the real world; the only purpose of these extreme assemblies is to show the possible potential of spatial impact depending on the distribution of coordinates. With other words, identical coefficients confirming *Zipf's* law may have a different meaning for different countries.

All simulated cases are then compared with respect to the stability of α and the significance of the spatial parameters ρ or λ respectively. I hypothesise that with a normally distributed assembly of coordinates and thus a zero level of spatial dependence, the spatial weight parameters are insignificant and meaningless, representing a *Zipf* distribution of the upper tail similar to that of words in a language. However, when modifying the coordinates and the spatial distribution of cities by building clusters of cities with a different level of size we may expect a stronger and significant impact of the spatial weight matrix. The interesting question is then how stable the original *Zipf* distribution of the upper tail remains.

Finally I examine the spatial influence on *Zipf's* law with Dutch, Slovene and Austrian natural cities extracted from night satellite images assuming that those better represent the true urban space as compared to a number of inhabitants within administratively defined city boundaries (background, segmentation methodology and data: cf. Bergs 2018; Bergs and Issa 2018; National Oceanic and Atmospheric Administration 2017).

3. Results

Table 1 shows the full rank-size distribution, while Table 2 displays the estimates only for the upper (*Pareto*) tail. First I take a look at the different results of the simulation exercise. Regarding case

I, *Zipf's* law is only confirmed for the curtailed distribution: α≈1 (Table 2). The lognormal extension of the smaller cities reduces the estimate of α to a large extent. As expected, the spatial weight coefficients λ of the full disribution and ρ of the *Pareto* tail are not significant. The normal distribution of coordinates of big and small cities leads to zero spatial dependence (*Moran's* I of the variable *Area* is close to zero). With or without the spatial extension of the model the α coefficient remains the same. *Zipf's* law is well confirmed for the *Pareto* tail.

Case II shows the same rank-size distribution but ordered geographically along the coordinates. There is a significant spatial influence confirmed by λ for the full distribution and ρ for the *Pareto* tail. The absolute value of α decreases substantially when incorporating spatial dependence. *Zipf's* law cannot be further confirmed for the upper (*Pareto*) tail, even though the random-generated distribution had been a *Pareto* (3, 1) one. The strong positive spatial autocorrelation of the variable *Area* (*Moran's* I=0.487) substantiates this finding.

For case III spatial autoregressive effects (ρ) are strong only with respect to the full distribution of cities. This is to be explained the spatial separation of big and small cities. Also here, spatial autocorrelation (*Moran's* I=0.429) of the variable *Area* is strong. In case of the big cities alone, there is no detectable spatial influence in the estimate. In both cases, the simple OLS regression as well as the extended spatial version *Zipf's* law can be confirmed (α=0.995 for both versions).

For Case IV (geographically ranked bigger cities, normally distributed small cities) spatial error effects (λ) are highly significant. However, α remains rather stable independent of whether or not incorporating spatial effects. For the full distribution the the absolute value of α increases by 0.03, for the upper tail it slumps from 0.995 to 0.986.

The simulated distributions of coordinates show that, in theory, spatial contiguity and distance may have a potential impact on the coefficient of the rank-size distribution of cities. The stylised models above are however rather artificial and not likely to be encountered somewhere in the real world. Therefore data of three EU countries are used to see how spatial effects may influence the estimate of rank-size distribution. As expected, the results generated are less spectacular than for the above cases II to IV but still suggesting spatial dependence to play a role in the rank-size distribution of cities for some countries.

In case of the Netherlands, estimated with 2011 radiance-calibrated DMSP-OLS night satellite images, a significant spatial influence can neither be established for the full distribution nor for the upper tail.

In case of Austria and Slovenia spatial spillover effects can be detected for the upper tail (bigger cities). Interestingly, spatial autocorrelation of the variable *Area* is negative, hence big and small cities are directly related via spatial distance. This would suggest a polycentric urban structure. For Austria, the spatial error estimate (λ) is significant at the p=0.05 level. But the influence as such is negligible; the α coefficient remains stable and slumps a little from 0.970 to 0.966. The Slovene case constitutes itself a bit different. Here the spatial autoregressive estimate (ρ) is also significant at the p=0.05 level, the α coefficient, however, changes from 0.983 (*Zipf*) to 0.860. This might have been implied by the small number of observations being in line with *Zipf's* law for the upper tail. It is to be noted that especially for small samples (as for small countries) the simple OLS estimate of *Zipf's* law is affected by underestimated standard errors. For all three countries I additionally estimated *Zipf's* law of the upper tail by using the *Gabaix-Ibragimov* approach to more precisely sort out the deviation when introducing spatial dependence (Gabaix and Ibragimov 2011):

$$\ln\left(R - \frac{1}{2}\right) = \ln(C) - \alpha \ln(S) + \varepsilon$$

The estimates are displayed by Table 3. While in case of Austria there is no difference in the estimate, there are major differences when it comes to Slovenia and the Netherlands. Regarding the former case α jumps from 0.983 to 1.178, thus beyond confirmation of *Zipf's* law; in the latter case it changes from 0.896 (slightly below the margin of *Zipf's* law) to 1.064 (well in the acceptable range of *Zipf's* law). There is no noticable change in standard errors.

## 4. Conclusion and further interpretation

Compared to *Zipf's* law for words in languages the results suggest that in case of cities, their spatial arrangement matters: *Zipf's* law for cities behaves like *Zipf's* law for words only if small and big cities are normally distributed in space. In theory, during time cities may change their size, but the slope of the rank-size distribution remains rather stable (Gabaix 1999; Duranton 2007). This is explained by its scale invariance and city growth independent of city size (*Gibrat's* law). Hence, the change of city ranks might be well explained by economic forces (structural change), but it is not directly visible in a changing slope of the rank-size distribution of cities. From a static point of view there is however reason to assume that specific location advantages and dispersion and concentration forces determine the spatial distribution of big and small cities. As soon as urban growth poles attract neighbour cities to grow, certain non-normally distributed patterns of spatial distribution of cities can evolve and spatial dependence of areal size increases (positive *Moran's I*). There is likewise the possibility of a more polycentric structure where big cities are functionally rather linked to smaller satellite towns. In this case, distance decay may outperform mass (size) and negative spatial dependence of areal size can be observed. If Gan et al. (2006) were right, and *Zipf's* law represents not more than a pure statistical relationship, the extension of the model with spatial effects (i.e. different functional relations between big cities and the smaller urban centers of rural areas) would not change α. Where such spatial impact is significant, *Zipf's* law for cities is more than a pure statistical phenomenon.

## References

Alexander C (1965) A City is not a Tree. Arch Forum 122(1): 58-62

Bergs R (2018) The detection of natural cities in the Netherlands – Nocturnal satellite inagery and Zipf's law. Rev Reg Res 38: 111-140

Bergs R, Issa M (2018) What do night satellite images and small-scale grid data tell us about functional changes in the rural-urban environment and the economy? – Case studies Frankfurt-Rhein-Main and Ljubljana Urban Region, PRAC, Bad Soden (Horizon2020 research grant 727988)

Brakman, S, Garretsen H, van Marrewijk C (1999), The return of Zipf: Towards a further understanding of the rank-size distribution. J Reg Sci 39(1): 183-213

Brakman S, Garretsen H, van Marrewijk C (2009), The New Introduction to Geographical Economics. Cambridge University Press, Cambridge

Chen Y (2015) The Distance-Decay Function of Geographical Gravity Model: Power Law or Exponential Law? Chaos Soliton Fract 77: 174-189

Cheng K, Zhuang Y (2012) Spatial Econometric Analysis of the Rank-size Rule for the Urban System: A Case of Prefectural-level Cities in China's Middle Area. Sci Geogr Sinica 32 (8): 905-912

Duranton G (2007) Urban Evolutions: The Fast, the Slow, and the Still. Am Econ Rev 97(1): 197-221

Eeckhout J (2004) Gibrat's Law for (All) Cities. Am Econ Rev 94(5): 1429-1451

Fujita M, Krugman P, Venables AJ (1999) The Spatial Economy: Cities Regions and International Trade. MIT Press, Cambridge (Mass.)

Gabaix X (1999) Zipf Law for Cities: An Explanation. Q J Econ 114(3): 739-767

Gabaix X, Ibragimov R (2011), Rank-1/2: A simple way to improve the OLS estimation of tail exponents. J Bus Econ Stat 29(1): 24-39

Gan L, Li D, Song S (2006) Is the Zipf law spurious in explaining city-size distributions? Econ Lett 92(2): 256-262

Jiang B, Miao Y (2014) The Evolution of Natural Cities from the Perspective of Location-Based Social Media. Prof Geogr 67(2): 295-306

Jiang B, Yin J, Liu Q (2015) Zipf's Law for all the Natural Cities around the World. Int J Geogr Inf Sci 29(3): 498-522

Krugman P (1997) Development, Geography and Economic Theory. MIT Press, Cambridge (Mass.)

Lalanne A (2014) Zipf's Law and Canadian Urban Growth. Urb Stud 51(8):1725-1740

Le Gallo J, Chasco C (2008) Spatial analysis of urban growth in Spain, 1900–2001. Empir Econ 34 59-80

Malevergne Y, Pisarenko V, Sornette D (2011) Testing the Pareto against the lognormal distributions with the uniformly most powerful unbiased test applied to the distribution of cities. Phys Rev E 83: 036111

National Oceanic and Atmospheric Administration (2017) Version 1 VIIRS Day/Night Band Nighttime Lights. https://ngdc.noaa.gov/eog/viirs/download_dnb_composites.html. Accessed 12 February 2018

Reggiani A, Nijkamp P (2015) Did Zipf anticipate Socio-economic Spatial Networks? Environ Plann B 42(3): 468-489

Rosen KT, Resnick M (1980) The Size and Distribution of Cities: An Examination of Pareto Law and Primacy. J Urban Econ 8(2): 165-186

Rybski D, García Cantú Ros A, Kropp JP (2013) Distance-weighted city growth. Phys Rev E 87:04214

Zipf GK (1949) Human Behaviour and the Principles of Least Effort. Addison Wesley, New York

**Annex**

|  | I | II | III | IV | The Netherlands 2011 (DMSP) | Slovenia 2017 (VIIRS) | Austria 2017 (VIIRS) |
|---|---|---|---|---|---|---|---|
| Endogenous variable: ln(Rank) | | | | | | | |
| ln(Area) | -0.403 | -0.322 | -0.321 | -0.432 | -0.424 | -0.739 | -0.833 |
| (Standard error) | (0.024)*** | (0.028)*** | (0.033)*** | (0.026)*** | (0.026)*** | (0.124)*** | (0.007)*** |
| Constant | 3.871 | 1.057 | 2.355 | 3.620 | 5.300 | 6.240 | 7.310 |
| (Standard error) | (0.037)*** | (2.998)*** | (0.435)*** | 0.483)*** | (0.787)*** | (0.742)*** | 0.120)*** |
| $\lambda$ | -0.321 | 0.969 | | 0.882 | - | - | 0.031 |
| (Standard error) | (0.650) | (0.030)*** | | (0.117)*** | - | - | (0.202) |
| $\rho$ | - | - | 0.392 | - | -0.159 | -0.110 | - |
| (Standard error) | - | - | (0.112)*** | - | (0.228) | (0.165) | - |
| Log Likelihood | -78.581 | -44.774 | -72.992 | -72.750 | -37.566 | 2.579 | 28.147 |
| Wald-Test $\lambda$ or $\rho$ = 0 ($\chi^2$) | 0.244 | 1016.678 | 12.293 | 57.052 | 0.484 | 0.441 | 0.023 |
| (p) | (0.622) | (0.000) | (0.000) | (0.000) | (0.487) | (0.507) | (0.879) |
| Moran's I (resid.) | 0.141 | 10.081 | - | 7.561 | - | - | 0.407 |
| (p) | (0.888) | (0.000) | - | (0.000) | - | - | (0.684) |
| Lagrange Multiplier (LM) | 0.163 | 80.748 | 14.883 | 19.539 | 0.310 | 0.436 | 0.027 |
| (p) | (0.686) | (0.000) | (0.000) | (0.000) | (0.578) | (0.509) | (0.870) |
| Robust LM | 0.028 | 15.462 | 11.544 | 39.497 | 0.260 | 0.671 | 0.042 |
| (p) | (0.867) | (0.000) | (0.001) | (0.000) | (0.610) | (0.413) | (0.838) |
| Moran's I (Area) | 0,014 | 0,473 | 0.429 | 0.123 | 0.053 | -0.016 | 0.010 |
| (p) | (0,039) | (0,000) | (0,000) | (0,000) | (0.043) | (0.095) | (0.038) |
| Obs. | 109 | 109 | 109 | 109 | 71 | 243 | 644 |
| | | | | | | | |
| OLS | | | | | | | |
| ln(Area) | -0.403 | -0.403 | -0.403 | -0.403 | -0.421 | -0.741 | -0.833 |
| (Standard error) | (0.025)*** | (0.025)*** | (0.025)*** | (0.025)*** | (0.026)*** | (0.123)*** | (0.008)*** |
| Constant | 3.870 | 3.870 | 3.870 | 3.870 | 4.757 | 5.748 | 7.310 |
| (Standard error) | (0.049)*** | (0.049)*** | (0.049)*** | (0.049)*** | (0.102)*** | (0.258)*** | (0.020)*** |
| Adj. $R^2$ | 0.709 | 0.709 | 0.709 | 0.709 | 0.789 | 0.937 | 0.944 |

Table 1: Spatial extension of Zipf's law (full rank-size distribution)

|  | I | II | III | IV | The Netherlands 2011 (DMSP) | Slovenia 2017 (VIIRS) | Austria 2017 (VIIRS) |
|---|---|---|---|---|---|---|---|
| Endogenous variable: ln(Rank) | | | | | | | |
| ln(Area) | -0.995 | -0.862 | -0.995 | -0.986 | -0.905 | -0.860 | -0.966 |
| (Standard error) | (0.017)*** | (0.029)*** | (0.018)*** | (0.019)*** | (0.070)*** | (0.171)*** | (0.022)*** |
| Constant | 5.086 | 3.846 | 5.117 | 4.937 | 7.202 | 6.540 | 7.482 |
| (Standard error) | (0.430)*** | (0.208)*** | (0.423)*** | (0.095)*** | (0.390)*** | (0.253)*** | (0.104)*** |
| $\lambda$ | - | - | - | 0.813 | -0.679 |  | -0.731 |
| (Standard error) | - | - | - | (0.180)*** | (0.919) |  | (0.345)* |
| $\rho$ | -0.056 | 0.262 | -0.067 | - | - | -0.550 | - |
| (Standard error) | (0.145) | (0.050)*** | (0.144) | - | - | (0.240)* | - |
| Log Likelihood | 39.618 | 50.343 | 39.652 | 42.830 | -2.167 | 8.389 | 28.731 |
| Wald-Test $\lambda$ or $\rho = 0$ ($\chi^2$) | 0.149 | 27.289 | 0.218 | 20.355 | 0.547 | 5.255 | 4.478 |
| (p) | (0.700) | (0.000) | (0.640) | (0.000) | (0.460) | (0.022) | (0.034) |
| Moran's I (resid.) | - | - | - | 5.072 | 0.148 | - | -1.531 |
| (p) | - | - | - | (0.000) | (0.882) | - | (1.874) |
| Lagrange Multiplier (LM) | 0.146 | 23.817 | 0.213 | 8.054 | 0.287 | 3.465 | 2.764 |
| (p) | (0.703) | (0.000) | (0.645) | (0.005) | (0.592) | (0.063) | (0.096) |
| Robust LM | 0.081 | 14.067 | 0.111 | 5.188 | 0.452 | 3.502 | 2.268 |
| (p) | (0.776) | (0.000) | (0.739) | (0.023) | (0.501) | (0.061) | (0.132) |
| Moran's I (Area) | -0.001 | 0.487 | 0.001 | 0.174 | 0.004 | -0.217 | -0.104 |
| (p) | (0.239) | (0.000) | (0.226) | (0.000) | (0.082) | (0.114) | (0.085) |
| Obs. | 50 | 50 | 50 | 50 | 21 | 11 | 50 |
| OLS | | | | | | | |
| ln(Area) | -0.995 | -0.995 | -0.995 | -0.995 | -0.896 | -0.983 | -0.970 |
| (Standard error) | (0.018)*** | (0.018)*** | (0.018)*** | (0.018)*** | (0.074)*** | (0.064)*** | (0.023)*** |
| Constant | 4.921 | 4.921 | 4.921 | 4.921 | 7.158 | 6.286 | 7.503 |
| (Standard error) | (0.039)*** | (0.039)*** | (0.039)*** | (0.039)*** | (0.421)*** | (0.310)*** | (0.109)*** |
| Adj. $R^2$ | 0.984 | 0.984 | 0.984 | 0.984 | 0.878 | 0.959 | 0.973 |

Table 2: Spatial extension of Zipf's law (rank-size distribution restricted on the upper tail)

|  | The Netherlands 2011 (DMSP) | Slovenia 2017 (VIIRS) | Austria 2017 (VIIRS) |
|---|---|---|---|
| Endogenous variable: ln(Rank-0.5) | | | |
| ln(Area) | -1.064 | -1.178 | -0.970 |
| (Standard error) | (0.071)*** | (0.074)*** | (0.023)*** |
| Constant | 7.994 | 6.918 | 7.003 |
| (Standard error) | (0.400)*** | (0.349)*** | (0.020)*** |
| Adj. $R^2$ | 0.922 | 0.965 | 0.974 |

Table 3: Zipf's law for the Netherlands. Slovenia and Austria (upper tail) estimated by the Gabaix-Ibragimov OLS approach

DRAFT 4 May 2020